\documentclass[a4paper,superscriptaddress,twocolumn,prl]{revtex4}
\usepackage{graphicx}
\usepackage[]{natbib}

\begin{document}

\title{A Bose-condensed, simultaneous dual species Mach-Zehnder atom interferometer}

\author{C.C.N. Kuhn}
\email{carlos.kuhn@anu.edu.au}
\homepage{http://atomlaser.anu.edu.au/}
\author{G.D. McDonald}
\author{K.S. Hardman}
\author{S. Bennetts}
\author{P.J. Everitt}
\author{P.A. Altin}
\author{J.E. Debs}
\author{J.D. Close}
\author{N.P. Robins}

\affiliation{Quantum Sensors and Atomlaser Lab, Department of Quantum Science, Australian National University, Canberra, 0200, Australia}

\date{\today} 

\begin{abstract}
This paper presents the first realisation of a simultaneous $^{87}$Rb -$^{85}$Rb Mach-Zehnder atom interferometer with Bose-condensed atoms.  A number of ambitious proposals for precise terrestrial and space based tests of the Weak Equivalence Principle rely on such a system. This implementation utilises hybrid magnetic-optical trapping to produce spatially overlapped condensates with a duty cycle of 20s. A horizontal optical waveguide with co-linear Bragg beamsplitters and mirrors is used to simultaneously address both isotopes in the interferometer. We observe a non-linear phase shift on a non-interacting $^{85}$Rb interferometer as a function of interferometer time, $T$, which we show arises from inter-isotope scattering with the co-incident $^{87}$Rb interferometer. A discussion of implications for future experiments is given.        
\end{abstract}

\maketitle

\section{Introduction}

Many decades of research have yet to unify the theory of gravity with the other fundamental interactions.  It is conceivable that gravity is simply a classical field \cite{PhysRevA.86.054101}, however most theorists have pursued a quantisation of gravity as the unifying theory \cite{Quantum_Gravity}.  Typically, these unifying theories imply a violation of General Relativity \cite{Tests_of_Lorentz_invariance}.  A number of ambitious experiments have been proposed to provide experimental data to test, and exclude or confirm, new theories.     

Atom interferometers provide an ideal platform for these experiments \cite{PhysRevLett.98.111102, MullerGravityTest, PhysRevD.68.124021, PhysRevA.86.033615}, providing ultra-high resolution absolute measurements of gravity   \cite{Schmidt}.  For example, simultaneously comparing a measurement of the acceleration due to gravity of two different atomic species allows a direct test of the weak equivalence principle of General Relativity. Indeed, large investments and spectacular progress have already been made in Europe and the US, most notably by the ESA projects Q-WEP and STE-QUEST in Europe \cite{Tino2013203, Sorrentino, Schubert_arxiv, Aguilera_arxiv}, and the Stanford drop tower project \cite{PhysRevLett.98.111102}.   The European team, in particular, have made outstanding contributions in terms of microgravity tests and the robustness of atom interferometry infrastructure \cite{PhysRevLett.110.093602, Geiger_airbourne}. A feature of the proposed experiments, both terrestrial and space based, is the use of a dual isotope $^{87}$Rb-$^{85}$Rb Bose-Einstein condensate (BEC) atom interferometer.  BEC is used to create a highly predictable, low expansion rate source \cite{RobinsAtomLaser} (with an equivalent classical gas temperature in the 10's of pK), a critical component of long interrogation time interferometry.  Additionally, use of the two isotopes suppresses common mode and platform noise by a predicted factor of $2.5\times10^{-9}$ \cite{Aguilera_arxiv}.  A very recent experiment demonstrated a simultaneous $^{87}$Rb-$^{85}$Rb atom interferometer based on a laser cooled source and Raman transitions, showing a factor of $1.8\times10^{-3}$ common-mode noise rejection \cite{PhysRevA.88.043615}.  

\begin{figure}[!hbp]
\centering{}
\includegraphics[width=1\columnwidth]{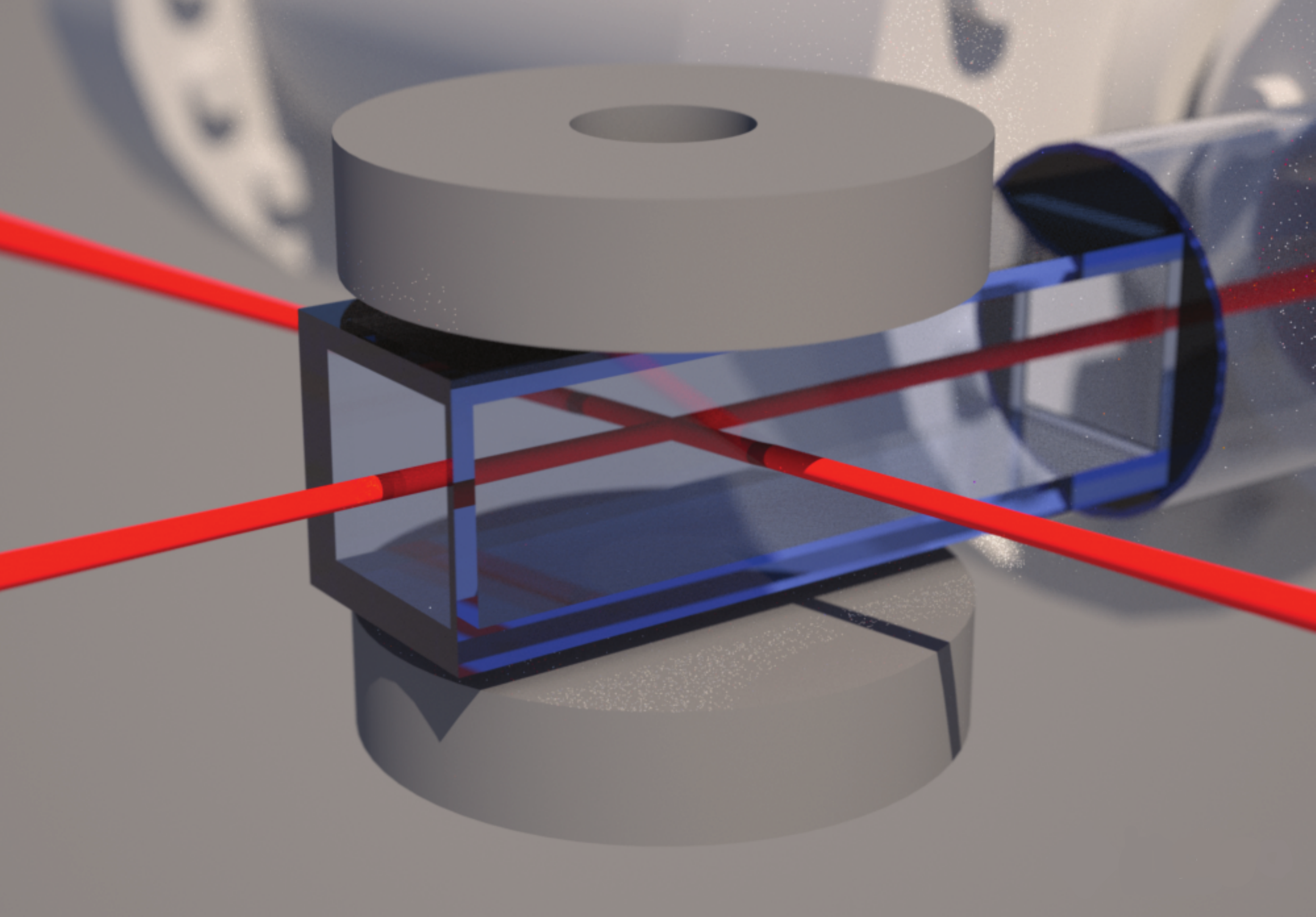}
\caption{(Color online)  A simplified schematic of the experiment, showing the key hardware for producing the dual condensates.  An ultra-high vacuum glass cuvette, quadrupole/feshbach coils and the crossed dipole trapping beams.  Not shown are the coincident two-colour MOT beams and the orthogonal absorption imaging systems. The glass cell is 3cm$\times$3cm$\times$10cm on its outside edges.}
\label{Set up diagram}
\end{figure}

This paper demonstrates the first simultaneous Mach-Zehnder atom interferometer based on Bose-condensed sources of $^{87}$Rb and $^{85}$Rb, confirming the broad methodology of both the European and US experiments.  The apparatus uses only two magnetic coils and a crossed dipole trap to produce variable number, simultaneous condensates (schematic shown in Figure 1). The BECs are then loaded into a horizontal optical waveguide, where an atom interferometer is constructed from co-linear counter propagating laser beams configured to drive Bragg transitions between momentum states.  It is shown that s-wave cross-scattering leads to a relative phase shift between the two isotopes. In addition to dual isotope interferometry for weak equivalence principle (WEP) tests, $^{85}$Rb offers a number of intriguing possibilities for atom interferometry.   The tuneable s-wave scattering of the $^{85}$Rb condensate allows control of the mean-field phase shift associated with condensate atom interferometers \cite{atomchip}, allowing interactions to be switched off.  This has been demonstrated in the context of lattice based Bloch oscillations \cite{PhysRevLett.100.080404,PhysRevLett.100.080405}. This work provides the first such demonstration in a Mach-Zehnder atom interferometer. Both the dual and non-interacting condensate interferometers provide important data for both future space missions and terrestrial high precision measurements.

This paper is organised as follows.  The apparatus is first described, focusing on the components and techniques that simplify production of $^{85}$Rb condensates. The tuneable nature of the $^{85}$Rb scattering length is demonstrated in both the crossed dipole trap and the waveguide.  Results are then presented for a simultaneous Bose-condensed $^{87}$Rb -$^{85}$Rb Mach-Zehnder atom interferometer and a non-interacting $^{85}$Rb atom interferometer.  The paper concludes with a discussion pertinent to the applications of this technique to proposed equivalence principle tests.       

\section{Apparatus}

The primary components of our experiment are a single set of low current ($\sim$15A) coils that can be dynamically configured to produce either a Helmholtz or quadruple configuration magnetic field, and two separate 20W laser beams that cross with large ($\sim$100$\mu$m) waists near the centre of the two coils.  This configuration is illustrated schematically in Figure 1.  This design, based on ideas from the $^{87}$Rb team at JQI  \cite{PhysRevA.86.063601, HybridTrap}, is a far simpler setup than our previous $^{85}$Rb apparatus \cite{altin:063103}.  Not surprisingly, this simplified system has many benefits, including much improved optical access and higher stability and repeatability in the condensate.         

\begin{figure}
\centering{}
\includegraphics[width=1\columnwidth]{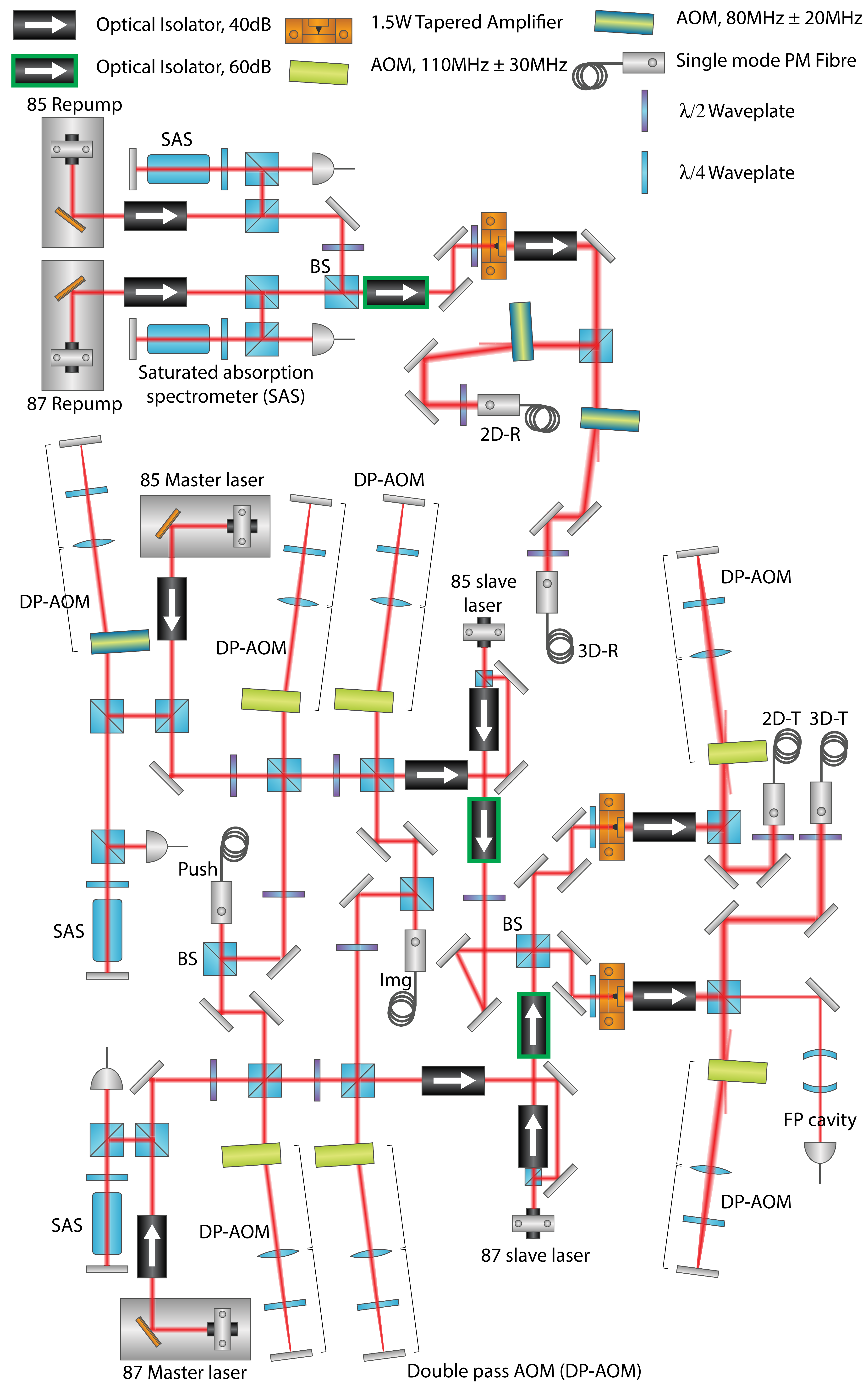}
\caption{(Color online) Schematic of the setup used to generate the lasers used in the experiment. The double pass AOM setup, sites the AOM and retro-reflection mirror a focal length away from the centre lens.  All beamsplitters are polarising except where noted as `BS', which denotes a non-polarising beamsplitter utilised for combining 87/85 laser beams with common polarisation.}
\label{laser_setup}
\end{figure}

\begin{figure*}[!htp]
\centering{}
\includegraphics[width=2\columnwidth]{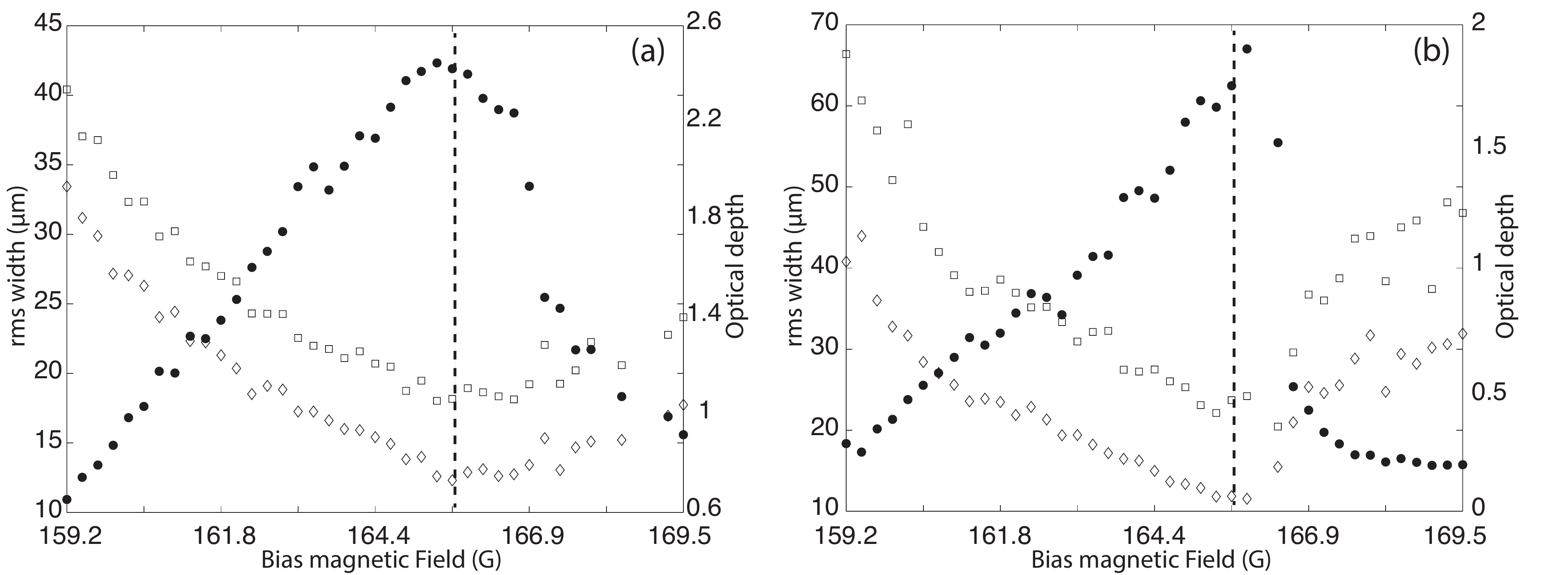}
\caption{(Colour online) (a) Expansion from the crossed dipole trap as a function of Feshbach magnetic field. Filled circles - OD, squares - $\sigma_x$, diamonds - $\sigma_y$. Vertical dashed line indicates zero scattering length on the horizontal axis. (b)  Expansion along the waveguide as a function of Feshbach magnetic field. Filled circles - OD, squares - $\sigma_x$, diamonds - $\sigma_y$.  Both sets of data represent a single sweep through the magnetic field values. }
\label{fringes}
\end{figure*}

The experiment is based around the same compact vacuum system described previously \cite{altin:063103}.  The basic optical design of the experiment has been retained as well, utilising a 2D-Magneto-Optical Trap (2DMOT) to load the primary 3DMOT via a pair of co-propagating push beams. Four different frequencies of light are required for laser cooling of the two rubidium isotopes.   The appropriate frequencies lie conveniently within a band of 6.8GHz at 350THz, easily accessible by readily available external cavity diode lasers.  Four separate lasers, one for each of the primary trapping and re-pumping transitions are used, each running at about 40mW, post isolator. This configuration has proved to provide the best stability for the experiment.  Sidebands added via diode modulation, external phase modulators or acousto-optic modulators have so far proved less stable than this configuration. By means of an AOM in the $^{85}$Rb master locking loop, both trapping lasers are locked with the same offset from the $|F=2\rangle \rightarrow |F' = 3\rangle$ and $|F=3\rangle \rightarrow |F' =4\rangle$ cooling transitions for $^{87}$Rb and $^{85}$Rb, respectively. The master lasers provide light for the push and imaging beams and for injection locking two separate free running diodes (running at 60mW each).  These injection locked lasers are mixed on a non-polarising beamsplitter and used to seed two tapered amplifiers (TAs).  By varying the alignment, mode matching and relative input powers, the output from the TA's are optimised to contain an appropriate ratio of the cooling light for two isotopes in the 2/3D MOTs.   Subsequent AOMs are used to rapidly vary the frequency and intensity of the light used for both MOTs, including shuttering.   Finally, all light is guided to the experiment head in high quality, single-mode, polarisation-maintaining optical fibers.  Repump light is produced in much the same way. Both repump lasers are locked to saturated absorption cross-over resonances with the nearly the same offset ($\sim$ 78MHz) from the re-pumping transitions $|F=1\rangle \rightarrow |F' = 2\rangle$ and $|F=2\rangle \rightarrow |F' =3\rangle$ for $^{87}$Rb and $^{85}$Rb, respectively.  The output from the repump lasers is mixed on a non-polarising beamsplitter and one port is used to inject a TA. Post-TA, the light is split through two separate single-pass AOMs and sent to the 2/3D MOTs.  Total trapping powers are 100mW and 120mW in the 3D and 2DMOTs, respectively, with approximately 15\% of that power at the $^{85}$Rb trapping frequency.  Repump powers are $\sim$15mW for each species in each MOT. The laser setup is designed to minimise non-common fluctuations between the cooling lasers, a critical factor in obtaining good stability in the final condensate atom numbers \cite{thesis_papp,thesis_altin,PhysRevLett.101.040402}.  The laser setup is shown schematically in Figure 2. 

The magnetic field for the 3DMOT, quadrupole magnetic trap and for accessing the $^{85}$Rb Feshbach resonance is generated by a single set of coils.  The coils are driven by a custom-built, closed-loop stabilisation circuit, providing a measured absolute RMS stability of better than 1:10$^6$ at a current of 15A (this corresponds to approximately 10$\mu$G stability around the Feshbach resonance). At all times, the current runs in series through the two coils. High speed switching is provided by an integrated capacitor boost circuit for the switch on (100$\mu$s from 0 to 15A), and TTL-controlled solid-state relays for the switch off (20$\mu$s from 15A to zero current).  An H-bridge switch electrically separates the two trapping coils, allowing a rapid switch of the direction of current in one of the coils (20$\mu$s). The pair of coils dissipate an absolute maximum of 40W total, and as a consequence no active cooling is required.       

The experimental sequence is separated into three stages: loading of the atoms in the 3D MOT, evaporative cooling in the hybrid trap, and the interferometer sequence.  
\subsection{3DMOT loading}
At the end of a 10s loading stage, $3\times10^6$ and $5\times10^8$ atoms of $^{85}$Rb  and $^{87}$Rb respectively have been collected.  Following 25ms of polarisation gradient cooling (PGC) on both isotopes (smoothly decreasing magnetic field and repump intensity while increasing MOT detuning), the repump light is extinguished and 1ms later also the trapping light, leaving a sample at a temperature of $\approx 15\mu$K.  This sequence pumps both isotopes into their respective ground states.  About 50\% of the atoms accumulate in our target states of $|F=1, m_F=-1\rangle$ for $^{87}$Rb and $|F=2, m_F=-2\rangle$ for $^{85}$Rb.  
\subsection{Dual species, hybrid trap cooling}
Immediately after PGC and optical pumping, the current in the trapping coils is switched to capture the atoms in a quadrupole trap, and then ramped to full current over 100ms, generating our maximum gradient of $\sim$300G/cm.  Simultaneously with the magnetic field ramp up, the power from two separate laser beams, intersecting at 30$\,^{\circ}$, is increased to produce a hybrid quadupole-magnetic and optical-dipole trap for the atoms. The axial confinement laser runs at $\lambda = 1090$ nm, with a $2$ nm linewidth, 25W power and waist 100$\mu$m.  The primary trapping laser (and waveguide) runs at $\lambda=1064$nm, with a $1$MHz linewidth, 20W power and waist 60$\mu$m. During $3.5$s, the magnetic and optical parameters of the hybrid trap are kept constant, while the clouds are cooled by radio frequency (rf) evaporation of $^{87}$Rb atoms in the $|F = 1, m_F = -1\rangle$ ground state. The $^{85}$Rb atoms in $|F=2, m_F=-2\rangle$ are cooled sympathetically, due to their tighter confinement \cite{PhysRevLett.101.040402,altin:063103}.   The magnetic field is then ramped to zero over 3s, loading the two isotopes into the pure crossed dipole trap, whilst forced rf evaporation continues until the magnetic field is off. At this stage we have a sample of $3.5\times10^{5}$ $^{85}$Rb and $2.5\times10^6$ of $^{87}$Rb atoms at temperatures close to $1\mu$K.

The H-bridge is then switched to the Helmholtz configuration, and the current is ramped up over 50ms to provide a bias field of $\sim$140G, and then rapidly jumped through the 155G Feshbach resonance to 165.64G, zeroing the s-wave scattering of $^{85}$Rb and minimising three-body loss \cite{PhysRevLett.85.728,PhysRevA.81.012713}.  Over a further $3.5$s the dipole trap intensities are smoothly ramped down to their final value giving an optical trap with radial and axial trapping frequencies of 70Hz and 9Hz, respectively. In the last $0.5$s of this ramp the bias field is tuned to give a $^{85}$Rb s-wave scattering length of 300$a_0$.  In this way, pure BECs of $2\times10^{4}$ $^{85}$Rb and $2\times10^{5}$ $^{87}$Rb are formed.  Depending on the details of the 3DMOT loading sequence, the condensate isotope ratio can be precisely controlled, from pure $^{87}$Rb, through a mixture, to pure $^{85}$Rb.  

After creation of the dual species condensate, the $^{85}$Rb s-wave scattering length can be tuned by varying the current in the Fechbach coils.  Indeed, this tuneability is a key feature to reduce systematics in the latest space interferometer proposals \cite{Schubert_arxiv, Aguilera_arxiv}.  Figure 3(a) demonstrates the tuneable nature of the s-wave scattering length in $^{85}$Rb.  Here, the condensate is created at 300$a_0$, and then the crossed dipole trap is rapidly extinguished.  Simultaneously, the Feshbach magnetic field is jumped to different values, rapidly changing the expansion rate of the condensate.  After 15ms of free expansion, the Feshbach field is switched to zero and, after a further 10ms expansion, the atoms are imaged with absorption imaging. The observed change in width is a feature of a tuneable scattering length condensate, and does not appear for a thermal cloud \cite{PhysRevA.85.053647}.  
\subsection{Interferometer sequence}
To perform the interferometer sequence, the atom clouds are transferred into a single beam dipole trap, effectively forming a waveguide for the atoms \cite{PhysRevA.84.043618}.  The loading sequence follows our previous technique \cite{PhysRevA.88.053620}, with the addition that, after forming the $^{85}$Rb condensate, the scattering length is ramped to zero over 100ms, followed by a ramp in the power of the single beam dipole trap over 200ms.  The weaker cross beam is then switched off, releasing the clouds into the waveguide.  It is observed that tuning the s-wave scattering length of the $^{85}$Rb cloud to zero mitigates heating in this transfer.  Similarly to the release data from the crossed dipole trap, Figure 3(b) shows expansion of the $^{85}$Rb cloud in the waveguide.  Recently such an expansion was described as the production of a `bright soliton' in the region of small negative scattering length \cite{cornish}.
Additionally, RF sweeps, similar to those used to calibrate the Feshbach magnetic field \cite{altin:063103}, can be used to transfer each isotope into any magnetic state \cite{PhysRevA.81.012713, PhysRevA.87.033611}. This will allow measurement of the interferometric phase shifts due to external magnetic fields, their gradients, and cross-and-self-scattering between the different $m_F$ states of each isotope \cite{Rakonjac:12}. A systematic study is beyond the scope of this work, but is clearly an avenue for future experiments.       
\begin{figure}[!htp]
\centering{}
\includegraphics[width=1\columnwidth]{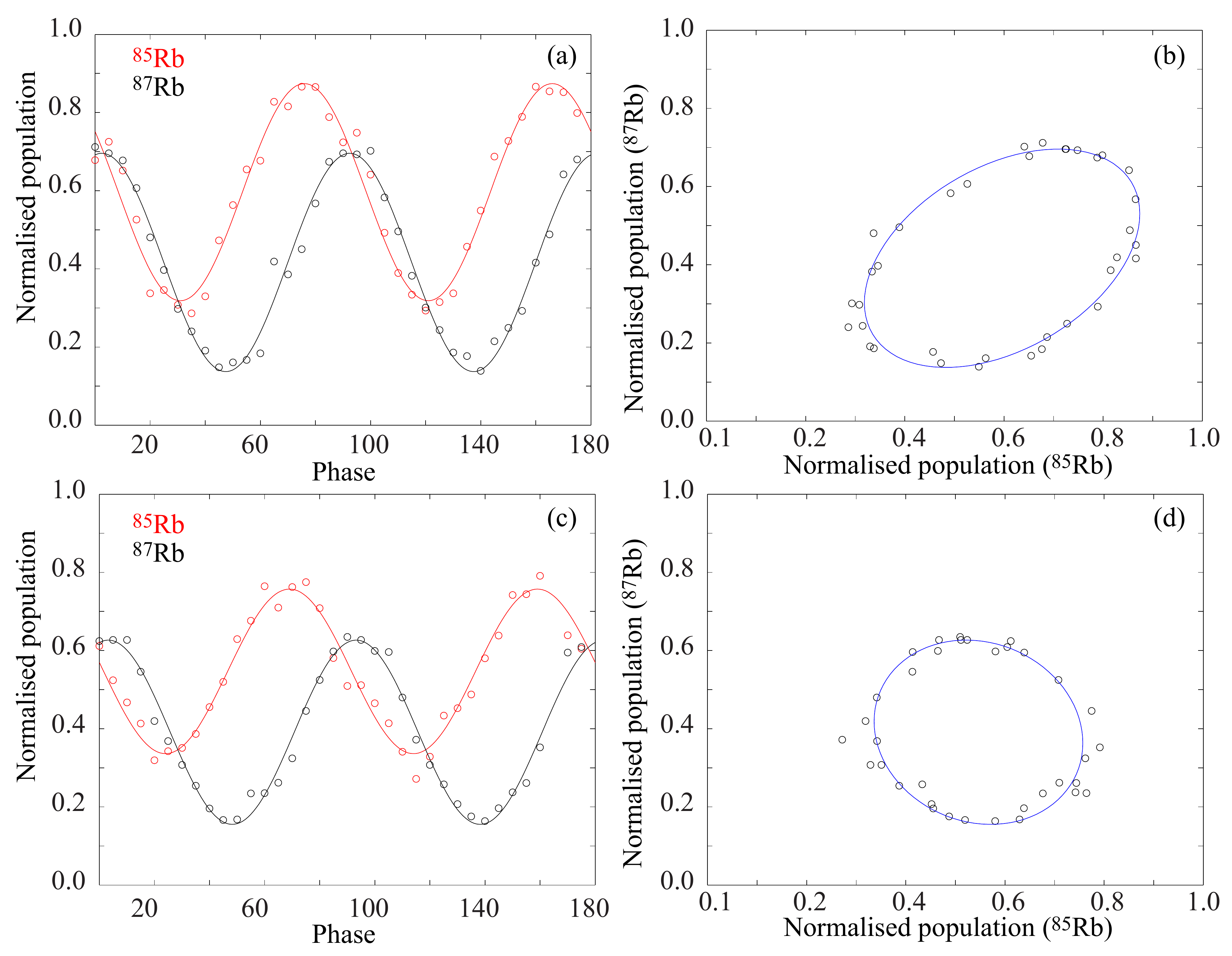}
\caption{(Color online)  (a) and (c) show the simultaneous fringes acquired from a 0.4ms and 0.5ms interferometer, respectively.  (b) and (d) show the ellipses generated from (a) and (c), respectively, by comparing normalised population in the two isotopes at varying phase.}
\label{fringes}
\end{figure}

After release into the waveguide, we engage a Bragg Mach-Zehnder interferometer.  The Bragg setup consists of up to 50 mW in each of two counterpropagating beams, precisely aligned to the waveguide optical trap by means of dichroic mirrors \cite{PhysRevA.87.013632}.  The lattice beams are collimated with a full width of 1.85 mm and detuned $\sim$105GHz to the blue from the D2 $|F=1\rangle \rightarrow |F' = 2\rangle$ and $|F=2\rangle \rightarrow |F' =3\rangle$ transitions of $^{87}$Rb and $^{85}$Rb, respectively. Arbitrary, independent control of the frequency, phase, detuning, and amplitude of each beam is achieved using a two-channel direct digital synthesiser (DDS) driving two separate AOMs.  The large detuning from both the $^{87}$Rb and $^{85}$Rb transitions sets the Bragg beamsplitter Rabi frequencies of the two isotopes equal to better than 1\%.  For this work, beamsplitters and mirrors with up to 8$\hbar$k of momentum transfer are used in a standard three pulse configuration.  Atom numbers in the momentum state output ports of the interferometer are measured using simultaneous absorption imaging of each isotopes on (near) orthogonal horizontal axes. 

A Mach-Zehnder Bragg interferometer is primarily sensitive to external accelerations \cite{OurGravimeter, PhysRevA.84.043618, PhysRevA.88.053620}, which lead to a phase shift on the interferometric fringes obtained.  This phase shift is the `signal' of the interferometer, which in our experiments is generated by a residual gravitational acceleration due to a small tilt in the horizontal waveguide.  Noise is generated by spurious accelerations of optical mirrors, vibration of the optical table and by possible fluctuating magnetic field gradients.   Figure 4 (a,c) shows typical interferometer fringes obtained by scanning the optical phase of the final Bragg beam-splitter using the DDS.  As observed in the laser-cooled simultaneous interferometer reported in \cite{PhysRevA.88.043615}, the fringes of each isotope have a relative phase shift, but unlike that work, the fringes are of very similar contrast.    This is because the Bragg laser system has only a single frequency in each counter-propagating beam, and hence drives the atoms with near identical Rabi frequencies. 
It is important to note that the ellipses generated by plotting the normalised atom number for each isotope against each other at each phase (Figure 4 (b,d)) indicate that there is non-common noise between the two interferometers.  It is most likely that this originates primarily from classical noise that is observed in our absorption imaging, and this is clearly a candidate for improvement in the apparatus.  Previously, we have experimentally observed shot noise limited interferometry with $10^4$ $^{87}$Rb \cite{PhysRevA.81.043633} atoms, and calculated that an observable interferometric signal-to-noise of 10,000:1 is possible with standard absorption imaging \cite{1367-2630-13-6-065020,1367-2630-13-11-119401}.
\begin{figure}
\centering{}
\includegraphics[width=1\columnwidth]{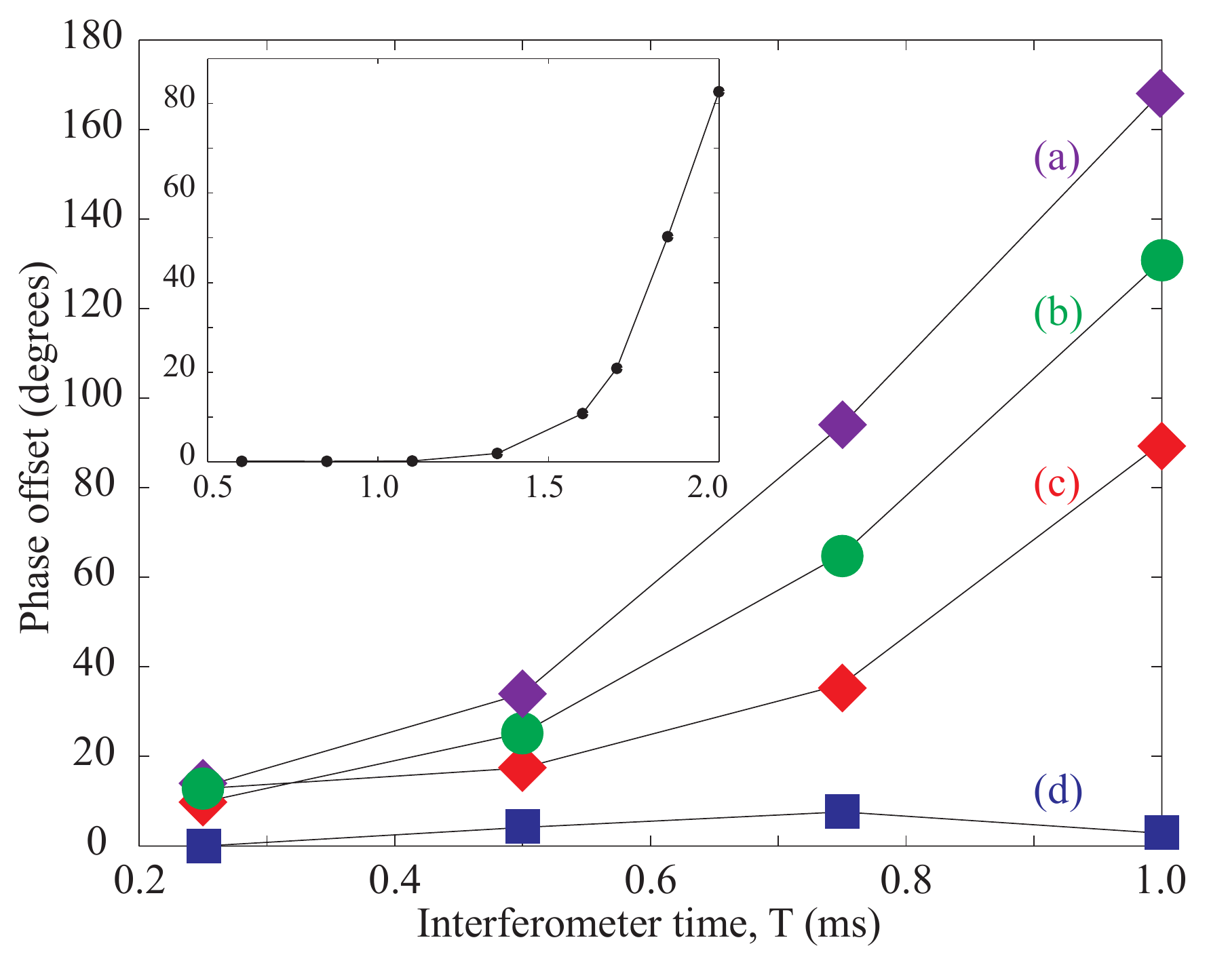}
\caption{(Color online)  Phase shift on $^{85}$Rb ($a_s\sim0$) fringes as a function of time, for varying atom number in the $^{87}$Rb co-incident interferometer. (a) $10^5$ atoms in $^{87}$Rb, (b) $6\times 10^4$ atoms in $^{87}$Rb, (c) $3\times 10^4$ atoms in $^{87}$Rb, (d) $\sim$0 atoms in $^{87}$Rb.  Inset: results of a numerical simulation based on a system of 1D Gross-Pitaevskii equations, showing qualitatively similar behaviour to the experiment for $N_{85}=2\times10^4$ and $N_{87}=1\times10^5$, $a_{85}=0$, $a_{87}=100a_0$ and $a_{85/87}=213a_0$.  A miscible initial condition in a 9Hz trap is used for this simulations. }
\label{}
\end{figure}

There is also a notable phase shift between the two isotopes in the data of Figure 4.  In order to investigate this relative shift, a $^{85}$Rb Bragg interferometer is operated at zero s-wave scattering length, and the shift studied as a function of the interferometer time with different numbers of $^{87}Rb$ present.  At $a_{85}=0$, numerical simulations indicate that there are no mean field dynamics in the pure $^{85}$Rb system, making it a near perfect interferometric sample \cite{Hardman_arxiv}.  With negligible $^{87}$Rb present, there is no obvious systematic phase shift on the $^{85}$Rb interferometer, as shown in Figure 5, curve (d).  However, as the number of atoms in $^{87}$Rb is increased there is a nonlinear increase in the phase shift of the $^{85}$Rb interferometer.  This strongly indicates that the relative phase shift, observed on the data of Figure 4 is due to the $a_{85/87}=213a_0$ inter-isotope s-wave scattering.  Reversing the wave-vectors of the Bragg beams, and operating a $-2\hbar k$ interferometer simply reverses the sign of this phase evolution, without changing its form.

A one-dimensional numerical simulation based on a system of coupled Gross-Pitaevskii equations \cite{Hardman_arxiv} for the two isotopes can be used to make a qualitative investigation of the observed effect.  The numerics confirm that there is a nonlinear phase shift due to presence of $^{87}$Rb, as shown in the inset to Figure 5.  There is no phase shift observed without $^{87}$Rb present. The numerical simulations indicate that there are complex spatial dynamics at play between the two isotopes. Quantitive agreement between experiment and theory may only be possible with a full three dimensional simulation.  Further study is needed to understand how the system will evolve over the long times needed for WEP tests.  This system can be considered as a weakly interacting two component spinor condensate, with potentially very complex initial conditions and dynamics \cite{RevModPhys.85.1191, PhysRevA.87.013625}.

\section{Discussion}
In this work, we have presented a robust and relatively simple arrangement to produce a dual-species Bose condensed atom interferometer. It is important to consider how our design and observations might contribute to discussions and planning on space-based systems.  

Current proposals use a magnetic chip trap to generate the fields for the MOT, and for creating an Ioffe-Pritchard (IP) configuration harmonic trap for evaporative cooling \cite{Schubert_arxiv, Aguilera_arxiv}.  Here we have shown that a more complicated macroscopic IP trap can be dispensed with altogether, in favour of a single pair of coils that provide the field for the MOT, magnetic trap and Feshbach magnetic field.  The configuration presented here may thus provide an alternative to a chip-based system, potentially providing greater optical access and simplicity than current designs.  It should be noted however, that the hybrid optical/magnetic potential used for pre-cooling would need to be carefully considered for operation in zero gravity.   

The number of condensed $^{85}$Rb atoms proposed for both terrestrial-and space-based atom interferometer experiments exceed current state of the art by nearly two orders of magnitude.  It is important to consider that to achieve fast and efficient sympathetic cooling of $^{85}$Rb by $^{87}$Rb in the pre-cooling stage requires a careful balance in atom number.  $^{85}$Rb will not efficiently evaporatively cool by itself due to an extremely low collision cross-section around 100$\mu$K, although it is feasible to produce condensates slowly \cite{PhysRevA.85.053647, PhysRevLett.85.1795}.  For the very large condensates being proposed, it is conceivable that lattice cooling \cite{PhysRevLett.89.090404,PhysRevLett.108.103001} will be required to overcome the stringent requirements placed on initial atom numbers by sympathetic cooling.   

Raman transitions are the mainstay of most atom interferometer precision measurement systems.  However, it was recently shown that the Bragg transitions used in this work are compatible with precision meansurement \cite{OurGravimeter}.   There are a number of interesting differences between Bragg transitions and the proposed Raman laser systems for space missions. For Bragg transitions, (i) only two counter-propagating laser beams, each with a single frequency, are required to drive the dual isotope interferometer, giving good common mode noise rejection on each isotope, (ii) there is no disadvantage with respect to internal states, as Bragg transitions can drive any internal state, (iii) detection via absorption imaging is simultaneous on all momentum states. Phase-contrast imaging, with the probe frequency set exactly between the two isotopes' imaging transitions, could potentially be used to directly extract a number difference giving near perfect imaging noise suppression (this technique could not be used in a dual isotope Raman interferometer), and finally (iv) Bragg transitions provide opportunities for large momentum transfer beam splitting to increase the machine sensitivity \cite{T3PRLArxiv}.  Clearly, enormous investments have been made in understanding systematic effects in Raman transitions, and it is in no way suggested that Bragg transitions provide a superior platform.  However, to provide an informed comparison, a study of systematic effects in high precision atom interferometry with Bragg transitions is required.     

This experimental technique, and simple vacuum system used here, could be straightforwardly retrofitted to other apparatus.  It is likely that dramatic improvements in laser power for CW cooling and optical beam splitting \cite{Sane:12} will also prove useful in terrestrial efforts.  Improvements to the duty cycle and large increases in collected atom number will likely lead to competitive measurements with terrestrial dual-condensate atom interferometers.

\section{Conclusion}

In summary, an apparatus to produce a dual-species BEC atom interferometer with a 20s duty cycle has been demonstrated.  A relative phase shift due to inter-isotope s-wave scattering was observed and numerically confirmed, although future 3D numerical studies will be required to gain a quantitative understanding of the system.  Our technique for condensate production is readily applicable, and perhaps critical, to ongoing efforts to produce experimental data in the realm of predicted violations of the universality of free fall.  

Avenues for further study include the production of a non-dispersive (soliton) atom interferometer, studies of $^{85}$Rb - $^{87}$Rb inter-isotope scattering and spinor dynamics, and an investigation of systematic effects in a high-precision Bragg atom interferometer.

\section{Acknowledgements}
The authors gratefully acknowledge the support of the Australian Research Council Discovery program. C.C.N. Kuhn would like to acknowledge financial support from CNPq (Conselho Nacional de Desenvolvimento Cientifico e Tecnologico). J.E. Debs would like to acknowledge financial support from the IC postdoctoral fellowship program.

\bibliographystyle{apsrev_v2}
\bibliography{dual_species_arxiv}
\end{document}